\documentclass[a4,12pt,epsf]{article}

\usepackage[dvips]{graphicx}
\begin{document}

\begin{flushright}
KANAZAWA-06-14\\
October, 2006
\end{flushright}

\begin{center}{\Large \bf
Anomaly of discrete family symmetries 
\vspace{6pt}\\
and gauge coupling unification 
\footnote{Talk given at the Summer Institute 2006, APCTP, Pohang, Korea.}
}
\end{center}

\begin{center}
Takeshi Araki
\footnote{araki@hep.s.kanazawa-u.ac.jp}
\vspace{6pt}\\

{\it 
Institute for Theoretical Physics, Kanazawa University, 
Kanazawa 920-1192, Japan 
}
\end{center}
\begin{abstract}
Anomaly of discrete symmetries can be defined as
the Jacobian of the path-integral measure.
We assume that an anomalous discrete symmetry
at low energy is remnant of an anomaly free discrete symmetry, 
and that its  anomaly
is cancelled by the Green-Schwarz (GS)
mechanism  at a more fundamental scale.
If the Kac-Moody levels $k_i$ assume non-trivial values,
the GS cancellation conditions of anomaly 
modify the ordinary unification of gauge couplings.
This is most welcome, because
 for a renormalizable model to be realistic
any non-abelian  family
symmetry, which should  not be  hardly broken {\em at low-energy},
requires multi $SU(2)_L$ doublet Higgs
fields.
As an example we consider a recently proposed
supersymmetric model with $Q_{6}$ family symmetry.
In this example, $k_2=1, k_3=3$ satisfies the GS conditions
and the gauge coupling unification appears
close to the Planck scale.
\end{abstract}

\section{Anomaly of discrete symmetries}
Anomaly is a violation of  symmetry at the quantum level.
In the case of a continuous symmetry, anomaly means non-conservation of 
the corresponding Noether current.
For  discrete symmetries, however,
there are no corresponding Noether currents.
But Fujikawa's method \cite{fujikawa},
which is based on the calculation of  
the Jacobian of the path-integral measure,
can be used to define anomaly of discrete symmetries.

Let us start by  considering a Yang-Mills theory with massless 
fermions $\psi$ in Euclidean space time,
which can be described by  the following Lagrangian and the path-integral:
\begin{eqnarray}
 &&Z=\int{\cal D}{\bar \psi}{\cal D}\psi
    \ {\bf exp}{\left[\int d^{4}x{\cal L}\right]},~
{\cal L}=i{\bar \psi}{\not D}\psi-\frac{1}{2g^{2}}
            {\bf Tr}\ F^{\mu\nu}F_{\mu\nu}, \\
 &&D_{\mu}=\partial_{\mu}-iA_{\mu}\ ,\ A_{\mu}\equiv gT^{a}A^{a}_{\mu}
   \ ,\ {\bf Tr}[T^{a}T^{b}]=\frac{1}{2}\delta^{ab},
\end{eqnarray}
where we have 
dropped the path-integral measure of the gauge boson $A_\mu$, 
because it does not contribute to anomaly.
Then we make a chiral phase rotation
\begin{eqnarray}
 \psi\rightarrow\psi^{'}={\bf e}^{i\theta\gamma_{5}}\psi, \label{eq,1}
\end{eqnarray}
where $\theta$ is a discrete parameter.
Under this finite transformation, the Lagrangian is invariant, but the
path-integral measure is not invariant in general, i.e.
\begin{eqnarray}
 &&{\cal D}\bar{\psi}{\cal D}\psi
           \rightarrow{\cal D}\bar{\psi}^{'}{\cal D}\psi^{'}
           =\frac{1}{J}{\cal D}\bar{\psi}{\cal D}\psi,
\end{eqnarray}
and the corresponding
Jacobian can be written as
\begin{eqnarray}
 &&J^{-1}={\bf exp}\left\{-i\int d^{4}x\ 
         \frac{\theta}{16\pi^{2}}{\bf Tr}
         \left[\epsilon^{\mu\nu\rho\sigma}
               F_{\mu\nu}F_{\rho\sigma}\right]\right\}.
\end{eqnarray}
This Jacobian has the same form as the Jacobian for a continuous 
transformation \cite{fujikawa}.
So we see that it makes sense to talk about 
anomaly of  discrete symmetries  \cite{IR,babu,kurosawa,dreiner}.

\section{The Green-Schwarz (GS) mechanism}

Unlike to \cite{babu}, in which only abelian discrete symmetries
are considered, we do not assume that the discrete symmetry
in question arises from the 
spontaneous break down of a  continues local symmetry.
We instead assume that an anomalous discrete symmetry
at low energy is remnant of an anomaly free discrete symmetry, 
and that its low energy anomaly
is cancelled by the GS mechanism \cite{GS} at a more fundamental scale.

First we discuss the abelian case and consider
a $Z_{N}$ transformation in a supersymmetric gauge theory,
\begin{eqnarray}
 \Phi\rightarrow {\bf e}^{-i\theta}\Phi,~~
 \bar{\Phi}\rightarrow \bar{\Phi}~{\bf e}^{i\theta},~~
 V\rightarrow V,
\label{eq,2}
\end{eqnarray}
where $\Phi$ and $V$ are a chiral superfield and a vector superfield,
respectively.
The transformation parameter
 $\theta$ is a discrete parameter, i.e. $\theta=\frac{2\pi}{N}$.
The Jacobian of this transformation appears in the superpotential
as \cite{konishi}
\begin{eqnarray}
 -i{\cal A}{\bf Tr}\left[\theta W^{a}W_{a}\right]_{F}
  +i{\cal A}{\bf Tr}
      \left[\theta\bar{W}_{\dot{a}}\bar{W}^{\dot{a}}\right]_{F},
\end{eqnarray}
where ${\cal A}$ is an anomaly coefficient, and $W$ is the chiral superfield
for the gauge supermultiplet.
This anomaly can be canceled by 
a shift of the dilaton superfield $S$
\begin{eqnarray}
 S\rightarrow S^{'}=S+i\frac{\cal A}{k}\theta,~~
 \bar{S}\rightarrow
 \bar{S}^{'}=\bar{S}-i\frac{\cal A}{k}\theta.
 \label{s}
\end{eqnarray}
where $k$ is the Kac-Moody level.
(For a non-abelian group, $k$ is a positive integer, while there is no
restriction in the abelian case.)
As we can see from
(\ref{s}), only the imaginary part of the scalar component of $S$,
which is the axion, should be shifted.
Therefore, the K\"{a}hler potential does not change, because
the vector superfield does not change under the transformation
(\ref{eq,2}).
From these observations, we can now
obtain the  anomaly cancellation conditions of the discrete symmetry
$Z_N$ \cite{babu}:
\begin{eqnarray}
 \frac{{\cal A}_{3}+\frac{pN}{2}}{k_{3}}
   =\frac{{\cal A}_{2}+\frac{qN}{2}}{k_{2}},\label{eq,3}
\end{eqnarray}
where ${\cal A}_{3}$ and ${\cal A}_{2}$ are anomaly coefficients of 
$[SU(3)_{C}]^{2}Z_N\ $ and $\ [SU(2)_{L}]^{2}Z_N\ $, respectively,
and  $p,q$ are integers.
The conditions containing products of
$U(1)_Y$ is omitted, because $k_{1}$ is not constrained to be an integer.
The constants $pN/2$ and $qN/2$ take into account
the  contributions from heavy Majorana and Dirac fermions
\cite{IR,babu,kurosawa,dreiner}.
As we will see in the next section,
Eq. (\ref{eq,3})  exhibits the anomaly cancellation conditions for non-abelian
discrete family symmetries, too.

\section{The GS mechanism for non-abelian 
 discrete family symmetries and 
unification of gauge couplings}
We first recall that the string coupling $g_{st}$ is the VEV of the dilaton field,
which is the real scalar component of the dilaton
superfield $S$.
Further, the gauge couplings $g_i$ are related to the string coupling 
according to 
the corresponding Kac-Moody levels.
Therefore,  gauge coupling unification conditions are
\begin{eqnarray}
 k_{3}g_{3}^{2}=k_{2}g_{2}^{2}=k_{1}g_{1}^{2}=g^{2}_{st}\label{eq,4}
\end{eqnarray}
at the string scale.
So, the unification conditions depend on the  Kac-Moody levels.
Keeping this in mind, we proceed with our discussion
on the non-abelian case.

Recently, a number of models with
a non-abelian discrete family symmetry
are proposed \cite{family}.
If only the SM Higgs is present within the framework
of renormalizable models, any non-abelian family
symmetry should be hardly broken.
That is, if a non-abelian family symmetry should be at most softly broken,
we need more than two $SU(2)_L$ doublet Higgs
fields.
This implies that the conditions of the ordinary unification of gauge
couplings, i.e. $k_2=k_3=k_1(3/5)$,  will be very difficult to be satisfied.
However, as  indicated in  Eq. (\ref{eq,4}), there is a possibility to
satisfy the unification conditions at the string scale for non-trivial
values of the Kac-Moody levels.
Before we study the unification conditions for a concrete model,
we derive the GS cancellation conditions for the non-abelian case below.
To this end, we consider the  Lagrangian
\begin{eqnarray}
 &&{\cal L}=i\bar{\psi}\not{D}\psi-\frac{1}{2g^{2}_L}
          {\bf Tr}\ F^{\mu\nu}(L)F_{\mu\nu}(L)
          -\frac{1}{2g^{2}_R}
          {\bf Tr}\ F^{\mu\nu}(R)F_{\mu\nu}(R),
   \\
 &&D_{\mu}=\partial_{\mu}-iL_{\mu}^{a}T^{a}P_{L}
                       -iR_{\mu}^{b}T^{b}P_{R},
\end{eqnarray}
and a non-abelian discrete chiral transformation
\begin{eqnarray}
 \psi_{\alpha^{'}}(x)\rightarrow\psi^{'}_{\alpha}=
   \left[{\bf e}^{iXP_{L}+iYP_{R}}\right]_{\alpha\beta}\psi_{\beta},
\end{eqnarray}
where $\alpha,\beta$ are family  indices.
Noticing that
this transformation is a unitary transformation, 
we then calculate the Jacobian
and find
\begin{eqnarray}
 J^{-1}={\bf exp}\left\{ i\int d^{4}x\ \frac{1}{32\pi^{2}}
       {\bf Tr}\ \epsilon^{\mu\nu\rho\sigma}\left[
         \theta_{L}\ F_{\mu\nu}(L)F_{\rho\sigma}(L)
            -\theta_{R}\ F_{\mu\nu}(R)F_{\rho\sigma}(R)\right]\right\},
\end{eqnarray}
where $\theta_{L,R}$ are defined as
\begin{eqnarray} 
\theta_{L(R)}\equiv{\bf Tr}[X(Y)],~~
{\bf e}^{i\theta_{L(R)}}\equiv{\bf det}\left[{\bf e}^{iX(Y)}\right].
\end{eqnarray}
Therefore, only the abelian parts of the non-abelian group 
contribute to anomaly,
implying that the GS   cancellation conditions for the non-abelian
case are exactly the same as Eq. (\ref{eq,3}) for the abelian case.

To be more concrete,
we consider the supersymmetric $Q_{6}$ model of \cite{Q6}.
According to our  discussion above,
to calculate anomaly of $Q_{6}$, it is sufficient
to consider anomaly of its abelian subgroups $Z_{6}$ and $Z_{4}$.
In the model of \cite{Q6}, it turns out that
the $Z_{6}$ part does not have any anomaly, but $Z_{4}$,
and its anomaly coefficients are computed as
\begin{eqnarray}
 &&2{\cal A}_{3}=2\cdot2-1-1=2\ (mod\ 4),~~
2{\cal A}_{2}=2\cdot3+2-1-1=6\ (mod\ 4).
\end{eqnarray}
This anomaly can be canceled by the GS mechanism, if
\begin{eqnarray}
\frac{1\ (mod\ 2)}{k_{2}}=\frac{3\ (mod\ 2)}{k_{2}}
 \label{level}
\end{eqnarray}
are satisfied.
In  string theory, 
lower Kac-Moody levels are preferable, and 
so $k_{2}=1,k_{3}=3$, for example, is a
preferable solution to (\ref{level}).
Let us see  whether 
the  unification conditions (\ref{eq,4}) 
with the levels satisfying (\ref{level})
can be satisfied.
To this end, we calculate the running of gauge couplings.
Fig.\ref{fig,1} shows the ratio of 
$g_2^2/g_1^2$ (upper line)
and $g_2^2/g_3^2$ (lower line) 
as a function energy scale.
From the figure we see that $g_2^2/g_3^2$
near the Planck scale $M_{PL}=1.2\times 10^{18}$ GeV is close to
$3$.
 Therefore, we assume that
the ratio of the two Kac-Moody levels is three, i.e.
$k_3/k_2=3$. 
Fig.\ref{fig,2} shows the running of 
$(\alpha_1 k_1)^{-1}, (\alpha_2 k_2)^{-1},
(\alpha_3 k_3)^{-1}$ with
$k_{3}=3, k_{2}=1$ and $ k_{1}\simeq1.63$.
The unification scale is 
$10^{20}$ GeV,
which is slightly higher than $M_{PL}$.
\begin{figure}[h]
 \begin{minipage}{.5\textwidth}
  \includegraphics[width=8cm,clip]{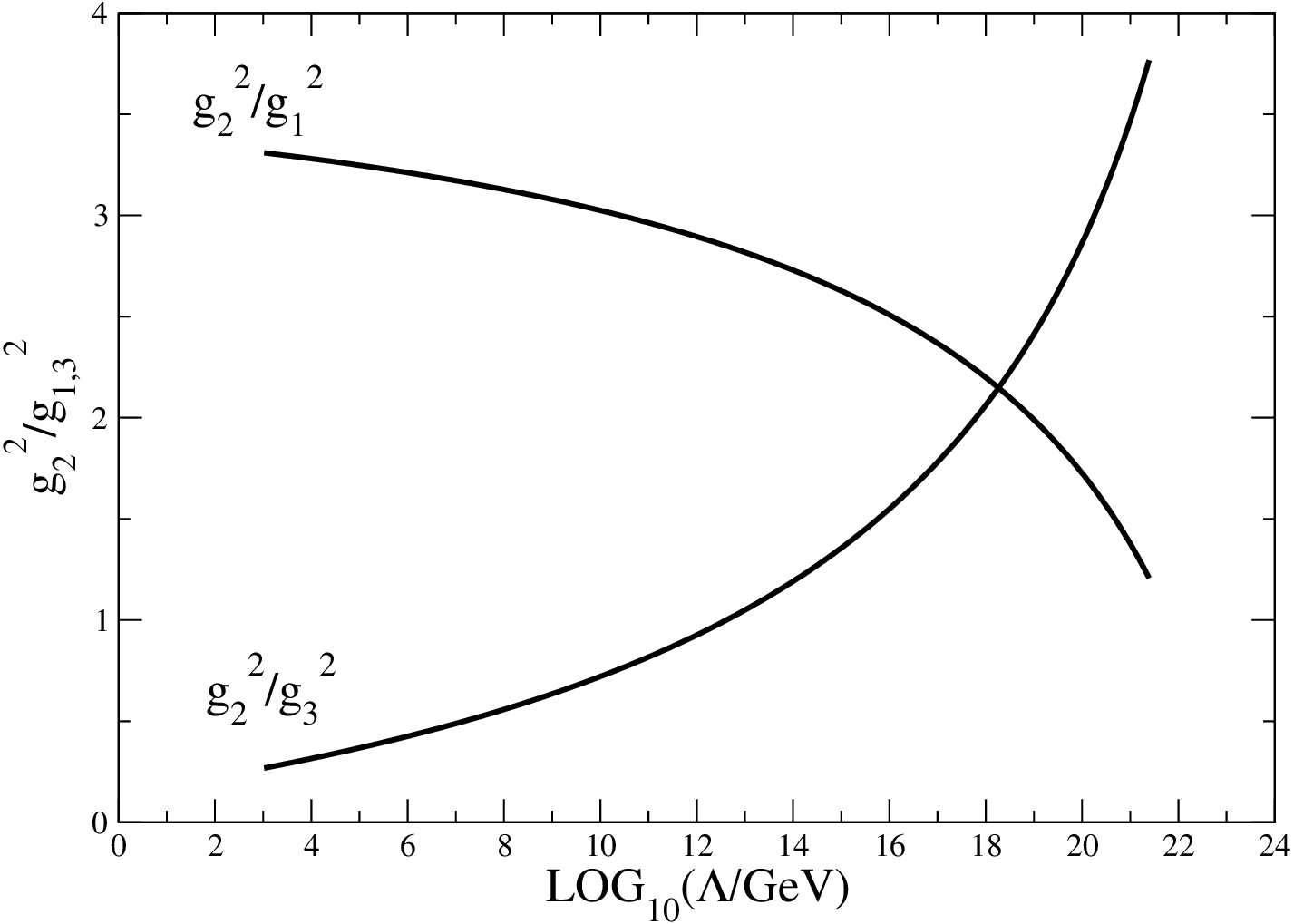}
  \caption{\footnotesize The ratio of
$g_2^2/g_1^2$ (upper line)
and $g_2^2/g_3^2$ (lower line) 
as a function energy scale.}
  \label{fig,1}
 \end{minipage}
 \begin{minipage}{.5\textwidth}
  \includegraphics[width=8cm,clip]{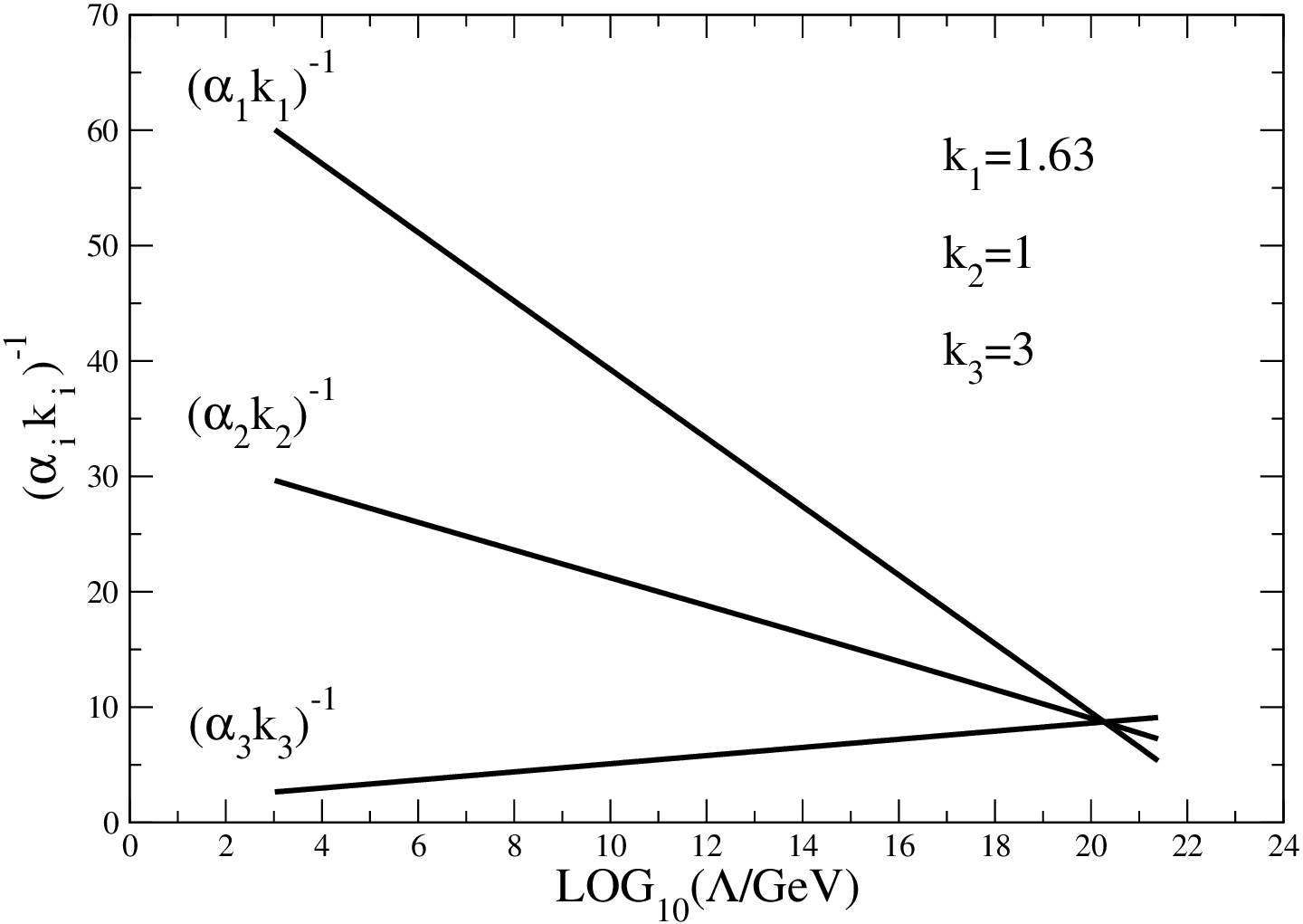}
  \caption{\footnotesize The running of 
$(\alpha_1 k_1)^{-1}, (\alpha_2 k_2)^{-1},
(\alpha_3 k_3)^{-1}$ with
$k_{3}=3, k_{2}=1$ and $ k_{1}\simeq1.63$.
The unification scale is 
$10^{20}$ GeV.
}
  \label{fig,2}
 \end{minipage}
\end{figure}
In this example the gauge couplings 
do not exactly satisfy the unification
conditions (\ref{eq,4}) at  $M_{PL}$, but it
is suggesting a right direction.
If  we  take into account 
the threshold corrections at  $M_{PL}$, for instance,
the conditions could be exactly satisfied.

Our study shows that
it is certainly worthwhile to look at  the other models more in detail.

\vspace*{12pt}
\noindent
{\bf Acknowledgement}
\vspace*{6pt}

\noindent
The Summer Institute 2006 is sponsored by the Asia 
Pacific Center for Theoretical Physics and the BK 21 
program of the Department of Physics, KAIST.
The author would like to thank the organizers of
Summer Institute 2006.


\end{document}